\documentclass[twocolumn,showpacs,aps,epsf,amsmath,amssymb,prb,superscriptaddress]{revtex4}
\usepackage{graphicx}
\usepackage{epstopdf}
\usepackage{amsbsy}
\usepackage[mathscr]{euscript}
\newlength{\onewidth}
\usepackage{tensor}
\usepackage{multirow}
\usepackage{epsfig}

\begin{document}

\title{Driving Perpendicular Heat Flow: Ambipolar Transverse Thermoelectrics for Microscale and Cryogenic Peltier Cooling
}

\author{Chuanle~Zhou}
\affiliation{
Electrical Engineering and Computer Science, Northwestern University, Evanston, IL 60208, USA
}
\author{S.~Birner}
\affiliation{
Walter Schottky Institut and Institute for Nanoelectronics, Technische Universit\"at M\"unchen, D-85748 Garching, Germany}
\affiliation{nextnano Semiconductor Software Solutions, M\"unchen, D-85586 Poing, Germany
}
\author{Yang Tang}
\affiliation{
Electrical Engineering and Computer Science, Northwestern University, Evanston, IL 60208, USA
}
\author{K. Heinselman}
\affiliation{
Electrical Engineering and Computer Science, Northwestern University, Evanston, IL 60208, USA
}
\author{M.~Grayson\footnote{Corresponding author: m-grayson@northwestern.edu}}
\affiliation{
Electrical Engineering and Computer Science, Northwestern University, Evanston, IL 60208, USA
}

\begin{abstract}
Whereas thermoelectric performance is normally limited by the figure of merit $ZT$, transverse thermoelectrics can achieve arbitrarily large temperature differences in a single leg even with inferior $ZT$ by being geometrically tapered. We introduce a band-engineered transverse thermoelectric with $p$-type Seebeck in one direction and $n$-type orthogonal, resulting in off-diagonal terms that drive heat flow transverse to electrical current. Such materials are advantageous for microscale devices and cryogenic temperatures -- exactly the regimes where standard longitudinal thermoelectrics fail.  InAs/GaSb type II superlattices are shown to have the appropriate band structure for use as a transverse thermoelectric.

\end{abstract}


\maketitle
In spite of their widespread success, conventional longitudinal thermoelectrics\cite{Rowe06} have limited use in microscale devices, at cryogenic temperatures, and over large thermal gradients, whereas transverse thermoelectrics have distinct advantages in all these regimes. Longitudinal thermoelectrics require multicomponent device structures with extrinsic p- and n-doped materials and multiple stages to achieve large thermal differences, while transverse thermoelectrics require only one single thermoelectric material, making microscaled devices straightforward. In addition, whereas longitudinal thermoelectrics suffer from dopant freeze-out at low temperatures, transverse thermoelectric phenomena can be optimal near intrinsic doping when both electrons and holes transport heat, making them operable at cryogenic temperatures. Exponentially tapered transverse thermoelectrics\cite{OBrien58} have demonstrated an additional advantage as infinite-stage Peltier refrigerators, predicted to cool to arbitrarily large temperature differences even with small figure of merit $ZT$, whereas longitudinal thermoelectrics require multiple stages\cite{OBrien} and large $ZT$ values.\cite{Rowe06}

Transverse thermoelectric phenomena require the directional symmetry of the Seebeck tensor to be broken. The Nernst-Ettingshausen (N-E) effect uses an external magnetic field to break time-reversal symmetry thereby introducing off-diagonal terms in the Seebeck tensor and generating transverse heat flow.\cite{Cuff} However practical application of the N-E effect is limited since a high 1.5~T magnetic field is required.\cite{Kooi} Stacked synthetic transverse thermoelectrics have also been demonstrated which have structural asymmetry by alternately stacking macroscopic millimeter-thick slabs of semiconductor with large Seebeck coefficient upon (semi)metal slabs with large electrical and thermal conductivity.\cite{Babin,Goldsmid10,Reitmaier,Mann,Kyarad05} Transverse heat flow is induced when the current is skewed with respect to the layers. However, these macroscopic extrinsic slabs do not permit microscale or cryogenic devices.

We develop a band engineering strategy for ambipolar transverse thermoelectrics (ATT), whereby the anisotropic electron and hole conductivity tensors give rise to a large transverse Seebeck coefficient in the absence of an external magnetic field. ATT materials can be geometrically shaped to enhance thermoelectric performance. Transport equations based on an electron-hole two-band model define the optimal angle of electric current for inducing the maximum transverse figure of merit $Z_\perp T$. Bulk compounds are identified which have the necessary ATT characteristics. To illustrate band engineering strategies, Type II broken gap InAs/GaSb superlattices (T2SL)\cite{Halasz}, which have been successfully used as infrared detectors\cite{Nguyen07}, are shown here to give promising $Z_\perp T$ values at room temperature for use as nanoscale transverse thermoelectric refrigerators. The equations for thermoelectric transport in an exponentially tapered cooler are also derived. The concept of a crossover electric field $\mathscr{E}^*$ is introduced to distinguish optimal performance in thin and thick samples.


\begin{figure}
	\includegraphics[width=60mm]{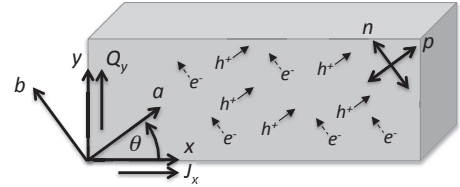}
    \caption{Ambipolar transverse thermoelectric has net $p$-Seebeck coefficient along $a$-axis with net $n$ along $b$, indicated with the crossed-arrow symbol on upper right. Microscopic electron-hole picture depicts net charge current $J_x$ to the right and net particle or heat current $Q_y$ up.}
	\label{fig:xyab}
\end{figure}

We begin with a description of how anisotropic electron and hole conduction gives rise to transverse thermoelectricity. Consider a semiconductor near intrinsic doping with anisotropic electron and hole band conductivity tensors $\boldsymbol{\sigma}_n, \boldsymbol{\sigma}_p$ and isotropic Seebeck tensors ${\bf s}_n, {\bf s}_p$, where $a$ and $b$ axes define the principle material axes of anisotropy:
\begin{eqnarray}
\label{eq:sigma}
\boldsymbol\sigma_n = \left[
 \begin{smallmatrix}
 \sigma_{n,aa} & 0 \\
 0 & \sigma_{n,bb}
 \end{smallmatrix}
 \right],
&
\boldsymbol\sigma_p = \left[
 \begin{smallmatrix}
 \sigma_{p,aa} & 0 \\
 0 & \sigma_{p,bb}
 \end{smallmatrix}
 \right]  \\
%
%
{\bf s}_n = \left[
 \begin{smallmatrix}
 s_{n} & 0 \\
 0 & s_{n}
 \end{smallmatrix}
 \right],
&
{\bf s}_p = \left[
 \begin{smallmatrix}
 s_{p} & 0 \\
 0 & s_{p}
 \end{smallmatrix}
 \right], \nonumber
\end{eqnarray}
where $s_n < 0$, $s_p > 0$. The total conductivity tensor $\boldsymbol{\Sigma}$ and total resistivity tensor $\boldsymbol{P}$ are $\boldsymbol{\Sigma} = \boldsymbol {P}^{-1} = \boldsymbol{\sigma}_n+\boldsymbol{\sigma}_p$.

The Seebeck tensor for these two parallel bands,
\begin{equation}
\label{eq:Seebeck_tot}
{\bf S} = (\boldsymbol{\sigma}_p + \boldsymbol{\sigma}_n)^{-1}
(\boldsymbol{\sigma}_p{\bf s}_p + \boldsymbol{\sigma}_n{\bf s}_n  )~,
\end{equation}
is fundamentally different from stacked synthetic transverse thermoelectrics\cite{Babin,Goldsmid10,Reitmaier,Mann,Kyarad05} whose out-of-plane Seebeck is set by series electrical and thermal resistances of two different materials. Whereas single-band Seebeck tensors ${\bf s}_n$ and ${\bf s}_p$ are typically isotropic, conductivity tensors $\boldsymbol\sigma_n$ and $\boldsymbol\sigma_p$ can be strongly anisotropic, and Eq.~\eqref{eq:Seebeck_tot} allows one to exploit this to weight the {\em total} Seebeck tensor to opposite signs for orthogonal directions. Defining a small parameter $\xi$ as the ambipolar conductivity ratio for a given direction, we define the $x$-direction as dominated by $p$-type conduction $\xi_a = \frac{\sigma_{n,aa} }{ \sigma_{p,aa}} < 1$, and the $y$-direction by $n$-type $\xi_b  =\frac{ \sigma_{p,bb} }{ \sigma_{n,bb}} < 1$. The variable $\xi$ determines how much the electron and hole contributions compensate the Seebeck effect in a given direction. The total Seebeck tensor
\begin{equation}
\label{eq:Peltier_shorthand}
{\bf S} = \left[
 \begin{smallmatrix}
 S_{p,aa} & 0 \\
 0 & S_{n,bb}
 \end{smallmatrix}
 \right]
\end{equation}
has matrix elements which are positive and negative, respectively, provided $\xi_a < |s_p / s_n|$ and $\xi_b <  |s_n / s_p|$,
\begin{eqnarray}
\label{eq:Peltier_pn}
S_{p,aa} =  \tfrac{s_{p} + \xi_a s_{n}}   { 1 + \xi_a } > 0,\\
S_{n,bb} =  \tfrac{s_{n} + \xi_b s_{p}}   { 1 + \xi_b } < 0. \nonumber
\end{eqnarray}

\begin{figure}[t]
	\includegraphics[width=85 mm]{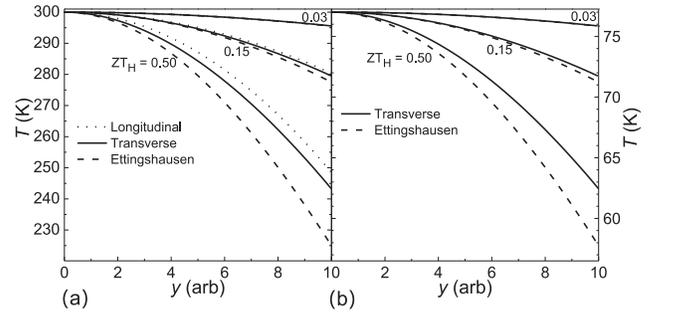}
    \caption{Optimal temperature profile for longitudinal (dotted line), transverse (solid line), and Ettingshausen (dashed line) thermoelectric cooling for various $ZT$ values assuming a Seebeck component $S_{xx}$ or $S_{xy}= 320 \frac{\mu \mathrm{V}}{\mathrm{K}}$, as appropriate. At $y = 10$ heat flow $Q_y = 0$, and at $y = 0$ heat sink is $T_{\mathrm H} = 300$\,K in panel (a), and $77$\,K in panel (b), where doped longitudinal thermoelectrics fail.}
	\label{fig:Zx3}
\end{figure}

As shown for stacked synthetic transverse thermoelectrics,\cite{Babin,Goldsmid10,Reitmaier,Mann,Kyarad05} such a diagonal tensor can yield off-diagonal Seebeck terms in a $\theta$-rotated $(x,y)$ transport basis, with current flow ${\bf J} = J_x \hat{\bf x}$ defining the $x$-axis [Fig.~\ref{fig:Zx3}~(a)]. Equations \eqref{eq:heatflow_trans}-\eqref{eq:ZTmax} apply in general to all transverse thermoelectrics,\cite{Babin,Goldsmid10,Reitmaier,Mann,Kyarad05}  but are rederived here to aid our subsequent discussion of tapered geometries.
With Peltier tensor $\boldsymbol{\Pi}$, the total Peltier heat flux density becomes ${\bf Q}_\Pi = \boldsymbol{\Pi}{\bf J} = (T{\bf S}){\bf J}$ with longitudinal and transverse components,
\begin{eqnarray}
\label{eq:heatflow_trans}
Q_{\Pi,x} = & {\bf Q} _\Pi \cdot \hat{\bf x} =& (S_{p,aa} \mathrm{cos}^2 \theta + S_{n,bb}  \mathrm{sin}^2 \theta  )\,T J_x \\
Q_{\Pi,y} = & {\bf Q} _\Pi \cdot \hat{\bf y} =& (S_{p,aa} - S_{n,bb})\, \mathrm{cos} \theta \mathrm{sin} \theta\,  T J_x.
\end{eqnarray}

The total heat flux density ${\bf Q} = {\bf Q}_\Pi - \boldsymbol{\kappa}^\mathrm{c} \boldsymbol{\nabla} T$ includes both Peltier and thermal conduction effects, where $\boldsymbol{\kappa}^\mathrm{c}$ as notated in Ref.\,\cite{Kooi} defines the open-circuit thermal conductivity tensor at ${\bf J} = 0$. Provided the thermal gradient is orthogonal to the current density $\boldsymbol{\nabla} T = \frac{dT}{dy} \hat{\bf y}$, the longitudinal electric field component $E_x$ is constant everywhere,\cite{Kooi} and the heat flux component $Q_y$ will depend only on $y$.
The longitudinal current and transverse heat flow are
\begin{eqnarray}
\label{eq:Ja}
J_x =     \tfrac{1}{\rho_{xx}}     E_x &-& \tfrac{S_{xy}}{\rho_{xx}}  \tfrac{dT}{dy} \\
\label{eq:Qb}
Q_y =  T \tfrac{S_{yx}}{\rho_{xx}}  E_x &-&(1+Z_{xy}T)\kappa^c_{yy}  \tfrac{dT}{dy},
\end{eqnarray}
with transverse figure of merit $Z_{xy}T = \frac{S_{xy}S_{yx}T}{\rho_{xx}\kappa^c_{yy}}$.
Steady state requires $\boldsymbol{\nabla} \cdot {\bf J} = 0$ and $\boldsymbol{\nabla} \cdot ({\bf Q} + \overline{\mu} {\bf J}) = 0$, where $\overline{\mu}$ is the electrochemical potential, and $-\boldsymbol{\nabla} \overline{\mu} = {\bf E}$ the electric field. Longitudinal Joule heating $E_x J_x$ sources a divergence in the transverse heat flux density $Q_y$:
\begin{equation}
\label{eq:div_Q}
\tfrac{dQ_y}{dy}=E_x J_x.
\end{equation}

Equations \eqref{eq:Ja}-\eqref{eq:div_Q} define the differential equation, 
\begin{eqnarray}
\label{eq:transcendental_temp}
0&= \frac{1}{S_{xy}S_{yx}} {E_x}^2
- \left [ \frac{S_{xy} + S_{yx}}{S_{xy}S_{yx}}
+ \frac{d\ln(S_{yx}/\rho_{xx})}{d\ln T} \frac{1}{S_{xy}} \right] E_x\frac{dT}{dy}+ \nonumber \\
&\left[ 1 + \frac{d\ln({S_{yx}S_{xy}/\rho_{xx}})}{d\ln T} + \frac{1}{Z_{xy}}\frac{d\ln \kappa^c_{yy}}{dT}    \right] \left( \frac{dT}{dy} \right)^2 \nonumber \\
&+ \frac{1+Z_{xy}T}{Z_{xy}} \frac{d^2T}{dy^2},
\end{eqnarray}
which with constant thermoelectric coefficients becomes
\begin{equation}
\label{eq:diffeq_temp}
0 =  \left(\tfrac{E_x}{S_{xy}}- \tfrac{dT}{dy}\right)^2+ \tfrac{1+Z_{xy}T}{Z_{xy}} \tfrac{d^2T}{dy^2}.
\end{equation}
Note this equation differs from the N-E effect, whose magnetic field requires $S_{xy} = -S_{yx}$ cancelling all $\frac{dT}{dy}$ terms in \eqref{eq:transcendental_temp} and permitting analytical integration, whereas for transverse thermoelectrics $S_{xy} = S_{yx}$, preserving the $\frac{dT}{dy}$ term and requiring numerical integration.

The angle $\theta_{\perp}$ which maximizes $Z_{xy} (\theta)T $ defines\cite{Babin,Goldsmid10,Reitmaier,Mann,Kyarad05} the parameter $Z_\perp T$
\begin{eqnarray}
\label{eq:thetaopt}
\cos^2 \theta_{\perp}  = &    \frac{1 }{1 + \sqrt{\frac{\kappa_{bb} / \kappa_{aa} }{\rho_{bb} / \rho_{aa}}}}  \\
\label{eq:ZTmax}
Z_{\perp}T = Z_{xy}(\theta_{\perp})T = & \frac{(S_{p,aa} - S_{n,bb})^2T}{({\sqrt{\rho_{aa}\kappa_{aa}} + \sqrt{\rho_{bb}\kappa_{bb}} })^2}.
\end{eqnarray}
The angle $\theta_{\perp}$ is independent of the Seebeck anisotropy, and approaches $\frac{\pi}{4}$ when the thermal conductance anisotropy matches the resistance anisotropy $\frac{\kappa_{bb} }{ \kappa_{aa}} = \frac{\rho_{bb} }{ \rho_{aa}}$. To optimize the electronic band structure it is useful to define a power factor $PF_{\perp}$ from Eq.~\eqref{eq:ZTmax} under the assumption of isotropic $\boldsymbol{\kappa}$ because ${\bf S}$ and ${\boldsymbol{\rho}}$ tensors can be readily calculated with simple scattering assumptions:
\begin{equation}
\label{eq:PF}
PF_\perp =  \tfrac{(S_{p,aa} - S_{n,bb})^2}{({\sqrt{\rho_{aa}} + \sqrt{\rho_{bb}} })^2}.
\end{equation}

The signature of ATT materials is thus the $p$-type Seebeck in one direction, and $n$-type perpendicular, arising from anisotropic band conductivity as shown above. Compounds have been shown to exhibit this property, with two noteworthy candidates RhGe$_x$Si$_{1.75-x}$ \cite{Gu} with $S_p = +160\,\mu$V/K and $S_n = -300\,\mu$V/K in the $a$- and $c$-directions, and CsBi$_4$Te$_6$ \cite{Chung} with $S_p = +100\,\mu$V/K and $S_n = -80\,\mu$V/K, in the $b$- and $c$-directions respectively. In addition, weaker ambipolar Seebeck anisotropies have been reported for PtCoO$_2$, PdCoO$_2$ and related materials.\cite{Ong} Although unipolar oxides like YBCO,\cite{He} LCMO,\cite{Zhao} and stoichiometries of CaCoO,\cite{Kanno} have been shown to exhibit a weak transverse Seebeck component $S_{ab} \sim 1-35~\mu$V/K when heat flows off-axis, the ambipolar effect introduced here can exceed this by an order of magnitude.

To illustrate how ATT behavior arises microscopically, we consider a band engineering example below. We introduce the InAs/GaSb T2SL\cite{Halasz} as a promising ATT with its tunable band gap and anisotropic electrical conductivity tensor. The bandgap is tunable down to zero gap since the GaSb valence band lies energetically {\em above} the InAs conduction band, thus the additional quantum confinement energy can tune the gap as the superlattice period is varied. The tunable gap has made the T2SL useful material for low-dark current infrared detectors and emitters.\cite{Nguyen07} The relatively large tunneling mass of holes and small tunneling mass of electrons makes the electrons to dominate the out-of-plane transport and with the appropriate chemical potential, the holes will dominate the in-plane transport. In the effective mass approximation, the conductivity and Seebeck tensor components become\cite{Chambers}
\begin{eqnarray}
\label{effective_mass_cond+Seebeck}
\sigma_{n,aa} &=& \tfrac{2 \sqrt{2} e^2 \gamma}{3 \pi^2 \hbar^3} \sqrt{m_{n,b}} (k_B T)^{s+\frac{3}{2}} \Gamma(s+\tfrac{5}{2}) F_{s+\frac{3}{2}}\left( \tfrac{\mu-E_g}{k_B T} \right) \nonumber \\
\sigma_{n,bb} &= &\tfrac{2 \sqrt{2} e^2 \gamma}{3 \pi^2 \hbar^3} \sqrt{\tfrac{m_{n,a}^2}{m_{n,b}}} (k_B T)^{s+\frac{3}{2}} \Gamma(s+\tfrac{5}{2})  F_{s+\frac{3}{2}}\left( \tfrac{\mu-E_g}{k_B T} \right) \nonumber \\
\sigma_{p,aa} &=& \tfrac{e^2\gamma}{\pi d\hbar^2} (k_B T)^{s+1} \Gamma(s+2) F_{s+1} \left(\tfrac{-\mu}{k_B T} \right) \nonumber \\
\sigma_{p,bb} &=& 0 \\
s_n &=& -\frac{k_B}{e} \left[\frac{(s+\frac{5}{2})F_{s+\frac{3}{2}}\left( \frac{\mu-E_g}{k_B T} \right)}{(s+\frac{3}{2})F_{s+1/2}\left( \frac{\mu-E_g}{k_B T} \right)} - \frac{\mu - E_g}{k_B T} \right] \nonumber \\
s_p &=& \frac{k_B}{e} \left[\frac{(s+2)F_{s+1}\left( \frac{-\mu}{k_B T} \right)}{(s+1)F_s\left( \frac{-\mu}{k_B T} \right)} + \frac{\mu}{k_B T} \right] \nonumber
\end{eqnarray}
where $m_i$ represents the effective mass of the band $n$ or $p$ in the direction $a$ or $b$. The electrochemical potential $\mu$ is measured relative to the valence band-edge, and $F$ is the Fermi-Dirac integral $F_r(\xi) = \int_0^\infty\xi^r \frac{1}{1+e^\xi}$. The scattering time $\tau = \gamma E^s$ obeys a power law in the kinetic energy of the carrier $E$. Since T2SL scattering at room temperature is dominated by interface scattering,\cite{Khoshakhlagh} we assume the power $s = 0$. Then $\gamma$ can be calculated from equation $\gamma = \tau =\frac{ \mu_c m^*}{e}$, $m^*$ where is the carrier effective mass and $e$ is the electron charge.

The band structure of the T2SL is calculated with the nextnano 8$\times 8 \,k \cdot p$ envelope function method.\cite{nextnano3,Grein} Applying Eq.\,\eqref{eq:PF} and Eq.~\eqref{effective_mass_cond+Seebeck}, the power factor $PF_{\perp}$ can be optimized over all possible $(N,M)$-SLs  where integers $N$ and $M$ count monolayers per period for InAs and GaSb, respectively. To minimize space-charge effects, we pin $\mu$ to the mid-gap, but note that p-doping can improve the performance. At 300\,K, the results yield (27,10)-SL as the optimal layer thicknesses. The electron and hole effective masses are $m_{n,a} = 0.028~m_0$, $m_{n,b} = 0.025~m_0$ and $m_{p,a} = 0.047\,m_0$ where $m_0$ is the free electron mass. The energy gap is $E_g = 34.2\,\mathrm{meV}$, with scattering coefficient $\gamma= 0.163$\,ps from experimental T2SL data.\cite{Khoshakhlagh} The longitudinal electrical resistivity is $\rho_{xx} = 255~\mathrm{m}\Omega$ and the transverse Seebeck coefficient is $S_{xy} = 320~\tfrac{\mu \mathrm{V}}{\mathrm{K}}$. The resulting optimal angle is $\theta_{\perp} = 37^\mathrm{o}$, which means that most of the heat flow will be perpendicular to the superlattice, so the thermal conductivity is dominated by the out-of-plane component, experimentally measured\cite{zhou} as $\kappa = 4~\mathrm{\tfrac{W}{m\cdot K}}$. The room temperature optimal figure of merit is $Z_{\perp} T = 0.030$. Typically out-of-plane thermal conductivity is less than in-plane $\frac{\kappa_{bb}}{\kappa_{aa}} < 1$, reducing the optimal angle $\theta_\perp$ even more according to Eq.\,\eqref{eq:thetaopt}.

Another recently developed type II superlattice material, the InAs/InAs$_{1-x}$Sb$_x$ superlattice, may also prove useful for ATT materials, having wider gaps and therefore operating at higher temperatures.\cite{Steenbergen} The ambipolar superlattices described here are to be contrasted with unipolar superlattice strategies which rely on quantum confinement \cite{Hicks} or topological interface states \cite{Tang} to enhance the unipolar Seebeck coefficient, but which do not consider transverse heat flow or simultaneous ambipolar contributions to the thermoelectric performance.

The maximum $\Delta T = T_\mathrm{H} - T_\mathrm{C}$ is reached when the heat load at the cold side is zero, $Q_{\mathrm{in}} (T_\mathrm{C}) = 0$. The temperature profile in Fig.~\ref{fig:Zx3}~(a) for this $Z_\perp T$ yields a total temperature difference $\Delta T$ = 4.5~K. To distinguish from other Peltier strategies, we compare to both longitudinal and N-E cooling for the same $ZT$ values and same Seebeck coefficient 320\,$\tfrac{\mu\mathrm{V}}{\mathrm{K}}$, and solve the temperature profile with $ZT_\mathrm{H}$ = 0.15 or $ZT_\mathrm{H}$ = 0.5 in Fig.~\ref{fig:Zx3}~(a). The maximum temperature differences of 20~K and 56~K, respectively, are shown. Transverse cooling (solid line) is slightly better than the longitudinal thermoelectric cooling (dotted line), and the thermal profile is distinctly different from the N-E effect for a hypothetical material with the same $ZT$ value (dashed line).

Because of their ambipolar nature, ATT can cool to arbitrarily low temperatures. Unlike standard longitudinal thermoelectric semiconductors which rely on extrinsic doping that freezes out at low temperatures, the thermoelectric cooling mechanism here is fundamentally intrinsic. Low operation temperatures for ATT can be achieved with sufficiently small $E_g$ such that electron and hole pairs can be thermally excited across the bandgap. Figure~\ref{fig:Zx3}~(b) shows the temperature profile with the heat sink temperature $T_\mathrm{H} = 77$~K for $ZT = 0.030, 0.15$ and 0.5, yielding the maximum temperature differences of 1.2~K, 5.3~K and 14.5~K, respectively.

\begin{figure}[b]
	\includegraphics[width=80 mm]{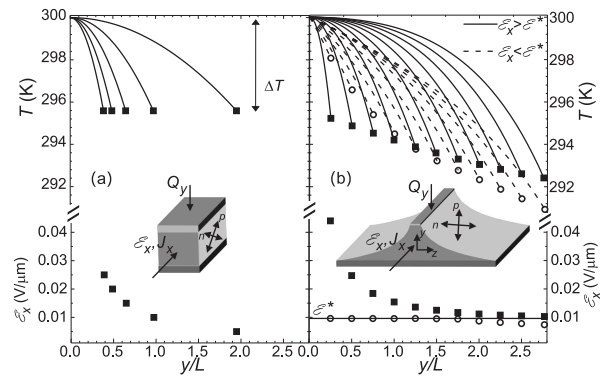}
    \caption{Temperature $T$ (top) and optimal electric field $\mathscr{E}_x$ (bottom) versus normalized sample thickness $y/L$, for rectangular (left) and exponentially tapered geometries (right). Exponential tapering constant $L$ sets length scale.  At $y = 0$ heat sink is $T_{\mathrm H} = 300$\,K and at right end point heat flow $Q_{\mathrm{in}} (T_{\mathrm{C}}) = 0$.  (a) For rectangular device, minimum $T_\mathrm{C}$ is independent of layer thickness, but thicker layers require weaker electric fields. Inset shows device geometry whereby current in $x$ direction induces heat flow in $-y$. (b) For exponentially tapered device, it can cool to lower temperatures with thicker material. Minimum $T_\mathrm{C}$ has two solutions for given sample thickness, one for $\mathscr{E}_x < \mathscr{E}^*$ and one for $\mathscr{E}_x > \mathscr{E}^*$.}
	\label{fig:rec_exp}
\end{figure}

More dramatic improvements in cooling power can be expected for transverse thermoelectrics by exploiting tapered geometries.\cite{OBrien58} As shown in Fig.~\ref{fig:rec_exp}~(b), an exponentially tapered device has a wide base on the heat-sink side, and a narrow strip at the refrigerated load. The exponential taper is in the $z$-direction, orthogonal to both the current direction and the heat flow, $z = z_0 e^{-y/L}$, and $L$ sets the characteristic length scale of the taper. Tapered N-E transverse coolers were shown to induce large temperature differences\cite{Kooi} when a tapered Bi$_{97}$Sb$_3$ semimetal in a 1.5~T magnetic field cooled from 156\,K down to 102\,K. Whereas cascaded longitudinal Peltier coolers are limited by electrical and thermal contact resistance, \cite{Rowe06} tapered transverse thermoelectrics result in infinite-stage cascading since the current and heat flow are perpendicular to each other,\cite{OBrien} allowing for much simpler device geometries and micron-scale fabrication.  Following the analysis of Kooi {\em et al.} for the N-E effect \cite{Kooi}, this exponential tapering adds the term of $Q_y \frac{d(\ln z)}{dy}$ to the right of Eq.\,\eqref{eq:div_Q}, and the following term to the right of Eq.\,\eqref{eq:diffeq_temp},  $- \left(\tfrac{1+Z_{xy}T}{Z_{xy}}\tfrac{dT}{dy}-\tfrac{E_x}{S_{xy}}T\right)\big/L.$
%
%

Figure\,\ref{fig:rec_exp} shows the T2SL temperature profile solved for $Z_\perp T = 0.030$ with different device thicknesses at optimal electric fields for both (a) a rectangular cooler without tapering, and (b) an exponentially tapered device. For the rectangular cooler, $\Delta T$ is independent of layer thickness, whereas the exponential tapering increases $\Delta T$ with increasing sample thickness. Assuming an experimentally proven tapering factor\cite{Scholz} of  $y/L = 2.77$, the ATT temperature difference should double to $\Delta T = 9.1$\,K. This is competitive with recent experimental results in on-chip cooling in significantly more complex device structures made of superlattice-based longitudinal thin-film thermoelectrics, which cooled only 7.1\,K on average.\cite{Chowdhury} For $Z_{\perp}T = 0.15$ and 0.5, the exponential tapering can also increase $\Delta T$ to 40 K and 97 K, respectively, almost doubling the temperature drop for all cases. For micron scale applications, the tapered structure could be a 10 $\mu$m thick ATT cooler with a 32 $\mu$m wide heat sink and a 2 $\mu$m wide cooled surface -- dimensions far smaller than that achievable with standard longitudinal coolers which require both $p$- and $n$-doped legs.

By setting $\tfrac{d^2T}{dy^2} = \tfrac{dT}{dy} = 0$ in the modified Eq.\,\eqref{eq:diffeq_temp}, the solution for $\mathscr{E}_x$ defines an important electric field scale for tapered devices, $\mathscr{E}^* = | \frac{S_{xy}T_\mathrm{H}}{L}|$.  When $\mathscr{E}_x  = \mathscr{E}^*$, ATT cooling perfectly compensates Joule heating and $T$ is everywhere constant. For $\mathscr{E}_x > \mathscr{E}^*$, $T(y)$ has a local maximum $\frac{dT}{dy}=0$ at $y=0$, and optimal thermal profiles for various sample thicknesses $y/L$ are shown with solid lines in Fig.\,\ref{fig:rec_exp}(b), top. For $\mathscr{E}_x < \mathscr{E}^*$, $\frac{dT}{dy}\ne0$ everywhere, and $\frac{dT}{dy}$ has a finite slope at $y=0$, plotted with dashed lines in Fig.\,\ref{fig:rec_exp}(b), top. In the bottom of panels Fig.\,\ref{fig:rec_exp}(a) and (b), the corresponding optimal electric fields are plotted, with a horizontal line indicating $\mathscr{E}^*$.  Large electric fields $\mathscr{E}_x > \mathscr{E}^*$ are seen to provide effective cooling for thin samples $y/L \lesssim 1$, whereas small electric fields $\mathscr{E}_x<\mathscr{E}^*$ cool better for thick samples $y/L \gtrsim 1$.

In conclusion, ATT materials will enable new regimes of thermoelectric operation in microscale and cryogenic devices. The intrinsic nature of the effect simplifies future thermoelectric devices, allowing a single layer of material to define a complete device, instead of traditional structures which require both $n$- and $p$-legs that become difficult to manufacture on the micron scale. One could envision, for example, a planar lithographic thin film of ATT to extract heat vertically from the surface upon applying an in-plane electrical current. The defining equations for the Seebeck tensor are shown to be distinctly different from previously studied stacked synthetic transverse thermoelectrics. Exponential tapering can enhance the thermoelectric performance, and even thin-film devices can be tapered to double the temperature difference. Thicker tapered materials will be able to cool to arbitrarily low temperatures even with small $Z_\perp T$. The intrinsic cooling mechanism can function at cryogenic temperatures, thus promising to fill a gap in thermoelectric cooling capabilities that currently exists below 150\,K.


This work was funded by the AFOSR grant FA9550-09-1-0237 and NSF-MRSEC grant DMR-0748856. MG would like to dedicate this manuscript to DCT.



\begin{references}






\bibitem{Rowe06} D.M. Rowe, {\it Thermoelectrics Handbook: Macro to Nano}, CRC Press (2006).

\bibitem{OBrien58} B. J. O'Brien and C. S. Wallace, J. Appl. Phys. {\bf 29}, 1010 (1958).

\bibitem{OBrien} B. J. O'Brien, C. S. Wallace, and K. Landecker, J. Appl. Phys. {\bf 27}, 820 (1956).

\bibitem{Cuff} K. F. Cuff, R. B. Horst, J. L. Weaver, S. R. Hawkins, C. F. Kooi, and G. M. Enslow, Appl. Phys. Lett. {\bf 2}, 145 (1963).

\bibitem{Kooi} C. F. Kooi, R. B. Horst, K. F. Cuff, and S. R. Hawkins, J. Appl. Phys. {\bf 34}, 1735 (1963).

\bibitem{Babin} V.P. Babin, T.S. Gudkin, Z.M. Dashevskii, L.D. Dudkin, E.K.Iordanishvilli, V.I. Kaidanov, N.V. Kolomoets, O.M. Narva, L.S. Stil'bans, Sov. Phys. Semicond. {\bf 8}, 478 (1974).

\bibitem{Goldsmid10} H. J. Goldsmid, J. Electron. Mater. {\bf 40} 5, 1254 (2010).

\bibitem{Reitmaier} C. Reitmaier, F. Walther, H. Lengfellner, Appl. Phys. A {\bf 99}, 717 (2010).

\bibitem{Mann} B. S. Mann, Master Thesis, Virginia Tech. (2006).

\bibitem{Kyarad05} A. Kyarad and H. Lengfellner, Appl. Phys. Lett. {\bf 87}, 182113 (2005); A. Kyarad and H. Lengfellner, Appl. Phys. Lett. {\bf 89}, 192103 (2006).

\bibitem{Halasz} G. A. Sai-Halasz, L. Esaki, and W. A. Harrison, Phys. Rev. B {\bf 18}, 2812 (1978).

\bibitem{Nguyen07}
B.-M. Nguyen, D. Hoffman, P.-Y. Delaunay, and M. Razeghi, Appl. Phys. Lett. \textbf{91}, 163511 (2007).

\bibitem{Gu}
J.-J. Gu, M.-W. Oh, H. Inui, and D. Zhang, Phys. Rev. B {\bf 71}, 113201 (2005).

\bibitem{Chung}
D.-Y. Chung, S. D. Mahanti, W. Chen, C. Uher, and M. G. Kanatzidis, Mat. Res. Soc. Symp. Proc. {\bf 793}, S6.1.1 (2004).

\bibitem{Ong}
K. P. Ong, D. J. Singh, and P. Wu, Phys. Rev. Lett. {\bf 104}, 176601 (2010).

\bibitem{He} Z. H. He, Z. G. Ma, Q. Y. Li, Y. Y. Luo, J. X. Zhang, R. L. Meng and C. W. Chu, Appl. Phys. Lett. {\bf 69} 3587, (1996)

\bibitem{Zhao} K. Zhao, K.-J. Jin, Y.-H. Huang, H.-B. Lu, M. He, Z.-H. Chen, Y.-L. Zhou, and G.-Z. Yang, Physica B {\bf 373} 72 (2006).

\bibitem{Kanno} T. Kanno, S. Yotsuhashi, and H. Adachi, Appl. Phys. Lett. {\bf 85}, 739 (2004); G. D. Tang, H. H. Guo, T. Yang, D. W. Zhang, X. N. Xu, L. Y. Wang, Z. H. Wang, H. H. Wen, Z. D. Zhang, and Y. W. Du, Appl. Phys. Lett. {\bf 98}, 202109 (2011).

\bibitem{Chambers}
After R. G. Chambers, "Electrons in Metals and Semiconductors," Chapters 9 and 10, Chapman and Hall (1990).


\bibitem{Khoshakhlagh} A. Khoshakhlagh, F. Jaeckel, C. Hains, J. B. Rodriguez, L. R. Dawson, K. Malloy, and S. Krishna, Appl. Phys. Lett. {\bf 97}, 051109 (2010).

\bibitem{nextnano3} next{\bf nano}$^3$ available at http://www.nextnano.de.

\bibitem{Grein} C.H. Grein, P.M. Young, M.E. Flatte, H. Ehrenreich, and J. Appl. Phys. {\bf 78} (12), 7143 (1995).


\bibitem{zhou} C. Zhou, B.-M. Nguyen, M. Razeghi, and M. Grayson, J. Electron. Mater., {\bf 41}, 2322 (2012).

\bibitem{Steenbergen}
E. H. Steenbergen, O. O. Cellek, D. Lubyshev, Y. Qiu, J. M. Fastenau, A. W. K. Liu, and Y.-H. Zhang, Proc. of SPIE {\bf 8268} 82680K (2012).

\bibitem{Hicks}
L. D. Hicks and M. S. Dresselhaus, Phys. Rev. B {\bf 47}, 16631 (1993).

\bibitem{Tang}
S. Tang and M. S. Dresselhaus, Nano Lett. {\bf 12},2021 (2012).

\bibitem{Scholz} K. Scholz, P. Jandl, U. Birkholz, and Z. M. Dashevskii, J. Appt. Phys. {\bf 75}, 5407 (1994).

\bibitem{Chowdhury} I. Chowdhury, R. Prasher, K. Lofgreen, G. Chrysler, S. Narasimhan, R. Mahajan, D. Koester, R. Alley, and R. Venkatasubramanian, Nat. Nanotechnol. {\bf 4}, 235 (2009).






\end{references}

\end{document}